# q-Gaussian Tsallis line shapes and Raman spectral bands

**Amelia Carolina Sparavigna**

Department of Applied Science and Technology, Polytechnic University of Turin, Italy

Email: amelia.sparavigna@polito.it

**Abstract**

q-Gaussians are probability distributions having their origin in the framework of Tsallis statistics. A continuous real parameter q is characterizing them so that, in the range 1 < q < 3, the q-functions pass from the usual Gaussian form, for q close to 1, to that of a heavy tailed distribution, at q close to 3. The value q=2 corresponds to the Cauchy-Lorentzian distribution. This behavior of q-Gaussian functions could be interesting for a specific application, that regarding the analysis of Raman spectra, where Lorentzian and Gaussian profiles are the line shapes most used to fit the spectral bands. Therefore, we will propose q-Gaussians with the aim of comparing the resulting fit analysis with data available in literature. As it will be clear from the discussion, this is a very sensitive issue. We will also provide a detailed discussion about Voigt and pseudo-Voigt functions and their role in the line shape modeling of Raman bands. We will show a successfully comparison of these functions with q-Gaussians. The role of q-Gaussians in EPR spectroscopy (Howarth et al., 2003), where the q-Gaussian is given as the "Tsallis lineshape function", will be reported. Two examples of fitting Raman D and G bands with q-Gaussians are proposed too. Torino, 19 March 2023.

**Keywords:** q-Gaussian distribution, Gaussian distribution, Cauchy distribution, Lorentzian distribution, Voigt distribution, Pseudo-Voigt function, Carbonaceous Materials, Raman spectroscopy, EPR spectroscopy, Tsallis line shape.

## 1. Introduction

q-Gaussians are probability distributions having their origin in the framework of Tsallis statistics (Tsallis, 1988, Hanel et al., 2009). The relevant functions of Tsallis statistics are the generalized forms of logarithm and exponential functions (see for instance the discussion in Sparavigna, 2022); a continuous real parameter q is characterizing them and when it is going to 1, the q-functions become the usual logarithm and exponential functions. The q-Gaussian is therefore the generalization of the Gaussian distribution. In the range 1 < q < 3, we pass form the Gaussian to a heavy tailed distribution. The value q=2, (Naudts, 2009), corresponds to the Cauchy distribution, also known in physics as the Lorentzian distribution. In heavy tail regions, the q-Gaussian is equivalent to the Student's t-distribution.

A change of the q parameter is therefore allowing the q-Gaussian to pass from Gaussian distribution to Lorentzian distribution. This property could be interesting for a specific application: that of a discussion about the shape of the Raman spectral lines.

The Raman bands are usually given as characterized by Lorentzian or Gaussian distributions, or by a l*inear combination* or by the *convolution* of them. In this last case, we have the Voigt distribution. Therefore, we will discuss q-Gaussians with the aim of comparing the resulting fit analysis with these distributions and data available in literature. As it will be clear from the discussion, in particular referring to literature (for instance Meier, 2005), this is a very sensitive issue. We will consider two specific cases too, regarding the analysis of D and G bands of carbon-based materials.

In Howarth et al., 2003, the q-Gaussian had been proposed as a line shape suitable for describing electron paramagnetic resonance (EPR) spectra, "and possibly nuclear magnetic resonance (NMR) spectra as well". In the article by Howarth et al., the q-Gaussian function is not mentioned in this manner, but as the "Tsallis lineshape". Howarth and coworkers stressed that the Tsallis line shape is generalizing the Gaussian and Lorentzian line shapes "widely used in simulations". The researchers compared the proposed Tsallis line shape with experimental EPR spectra. The q-line shape "often provides a better approximation of the experimental spectrum". It is also mentioned a link to intermolecular spin–spin interactions.

Before the discussion of q-Gaussian function and comparison, let us just give information about the Raman spectroscopy.

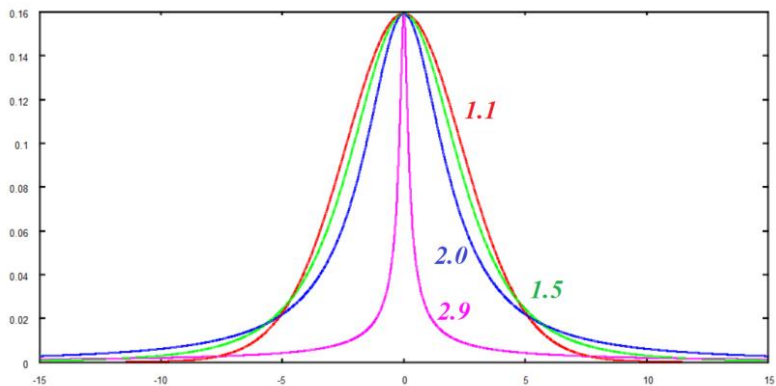

*Fig. 1: q-Gaussian functions, for different q indices, from 1.1 (quasi-Gaussian) to 2.9 (over-Lorentzian). The blue curve is the Lorentzian line shape.*

## 2. The Raman bands

The aim of this discussion is therefore to test the possibility of using q-Gaussians in another spectroscopy such as the Raman one. Raman spectroscopy is based on the interaction of light with a substance, evidencing chemical bonds in it. In this manner, information about chemical bonds, polymorphism and crystalline features can be deduced. The spectrum is characterized by bands, so that the related combinations are relevant for specific classes of materials. Several Raman band correlation tables exist, usually ranging from 100 $cm^{-1}$ to 3300 $cm^{-1}$. Then let us concentrate our discussion on a limited part of the spectrum, and the considered materials are the carbonaceous ones.

As told by Lünsdorf et al., 2017, the Raman spectrum of carbonaceous materials (CMs) is made of two major "intensity accumulations in the regions" of 1000–1500 $cm^{-1}$ and 1500–1700 $cm^{-1}$; these regions are given by Lünsdorf and co-workers as D- and G-band regions. "The intensity distribution in these regions depends on the maturity stage or degree of graphitization and used excitation wavelength" (Lünsdorf et al., 2017, and references therein). For the shapes of these regions, see please the Figure 4 of the given reference. In the upper part of the figure, "none graphitized CM", we can see how the spectrum has been fitted. Lünsdorf and coworkers say that these regions are "usually curve-fitted by five Lorentz functions (Lahfid et al., 2010) or a mixture of four Lorentz and one Gauss functions (Sadezky et al., 2005). For the graphitic CM, we can see that the regions reduced to three bands (see the lower part of the Figure 4 in Lünsdorf et al., 2017). In this case, each band is usually described by a Voigt function. Lünsdorf and coworkers use a software that simultaneously evaluates the background and the Raman signal, respectively simulated by a fifth-order polynomial and pseudo-Voigt functions.

In Sadezky et al., 2005, we find experiments and fitting procedures for the Raman spectra of carbonaceous materials too. The aim of researchers was that of optimizing a Raman microscope system using three different laser excitation wavelengths (514, 633, and 780 nm). The band combinations involved four Lorentzian-shaped bands (G, D1, D2, D4) at about 1580, 1350, 1620, and 1200 $cm^{-1}$, and a Gaussian-shaped band (D3) at about 1500 $cm^{-1}$ in the framework of first-order spectra. The second-order spectra were fitted considering Lorentzian-shaped bands at about 2450, 2700, 2900, and 3100 cm$^{-1}$. In the Table 1 of Sadezky et al., 2005, first-order Raman bands and vibrational modes in soot and graphite are given as follow.

| Band | Raman-shift | | | Vibration mode |
|---|---|---|---|---|
| | Soot | Disordered graphite | High-ordered graphite | |
| G | ~ 1580 | ~ 1580 | ~ 1580 | Ideal graphitic lattice |
| D1 | ~ 1350 | ~ 1350 | - | Disordered graphitic lattice |
| D2 | ~ 1620 | ~ 1620 | - | Disordered graphitic lattice |
| D3 | ~ 1500 | - | - | Amorphous carbon (Gaussian or Lorentzian line shape) |
| D4 | ~ 1200 | - | - | Disordered graphitic lattice |
| | (cm$^{-1}$) | | | (Lorentzian line-shape unless mentioned otherwise) |

Sadezky and coworkers approached the fitting in the following manner. They stated nine different combinations for the first-order Raman bands. The combinations are given in their Table 2, where we can find the used line shapes (L Lorentzian or G Gaussian), and also initial band positions. The nine combinations are as in the following table. The best result was obtained with the combination IX; as told before the combination was of Lorentzian-shaped bands G, D1, D2, and D4 and Gaussian-shaped band D3. That is, Sadezky and coworkers used only line shapes of Lorentzian and Gaussian forms.

| Band | I | II | III | IV | V | VI | VII | VIII | IX |
|---|---|---|---|---|---|---|---|---|---|
| G | L | L | L | L | L | L | L | L | L |
| D1 | L | L | L | L | L | L | L | L | L |
| D2 | - | - | - | L | - | - | L | L | L |
| D3 | L | G | - | - | L | G | - | L | G |
| D4 | - | - | L | - | L | L | L | L | L |

A discussion about Raman spectroscopy for investigating the role of thermochemical processes of coal and biomass is given by Xu et al., 2020. Xu and coworkers consider the Raman spectroscopy as a versatile tool for studying biomass. In the Table 2 of the given reference, we can find the description of the Raman bands for samples from graphite to amorphous carbon.

We see the D4 band at ~ 1200 cm$^{-1}$, D1 ~1350 cm$^{-1}$, D3 ~1540 cm$^{-1}$, G ~1590 cm$^{-1}$, and D2 at ~1620 cm$^{-1}$.

In the Table 3 we can find also the curve-fitting methods for the first-order Raman spectra as:

1) two bands with components about 1350 (D1), 1590 (G);

2) three bands with components about 1350 (D1), 1500 (D3), 1590 (G), or about 1350 (D1), 1590 (G), 1620 (D2), or 1200 (D4), 1350 (D1), 1590 (G);

3) four bands with components about 1200 (D4), 1350 (D1), 1500 (D3), 1590 (G);

4) five bands about 1200 (D4), 1350 (D1), 1500 (D3), 1590 (G), 1620 (D2), or about 1080, 1200 (D4), 1350 (D1), 1500 (D3), 1590 (G); or 1200 (D4), 1350 (D1), 1500 (D3), 1585 (G), 1750.

And more bands are also mentioned. Xu et al., 2020, are giving related references for the models given in their Table 3. Among the five bands method, Xu and coworkers are mentioning Sadezky et al. (in the Figure 7 of Sadezky and coworkers' article, we can find the fit with band combination IX for the first-order Raman spectra of diesel soot).

As an example of four-band fitting, Xu et al. are referring the work by Han et al., 2017, about the role of structure defects in anthracite. The Fig. 4 of this reference is showing the Raman spectra of collected samples (a) and the fitting on one of them in (b); the fit is made by means of bands G, D, D3 and D4. Two major bands respectively at 1350 and 1600 cm$^{-1}$ are identified. A minor band at about 1230 cm$^{-1}$ is also visible. The researchers used PeakFit, version 4.12, software (Han et al., 2017). The resulting four bands are: G band (1597 cm$^{-1}$, Lorentzian profile), D band (1338 cm$^{-1}$, Lorentzian profile), D3 band (1499 cm$^{-1}$, Gaussian profile), and D4 band (1237 cm$^{-1}$). A discussion about the corresponding vibrational modes is given too.

In the analysis process of Raman spectra, components are added to improve "the accuracy of the fit", as told by Sousa et al., 2020, who studied the structural alteration of biochar. Let us just tell that biochar is the solid residue of pyrolysis of biomass, obtained by thermochemical decomposition at moderate temperatures under oxygen-limiting conditions (Brassard et al., 2019). The principal use of biochar is as amendment in agricultural soils, (Brassard et al., 2019). For other applications, such as in Shape-Stabilized Phase-Change materials, see please Sparavigna, 2022, and references therein (see also Bartoli et al., 2020, 2023).

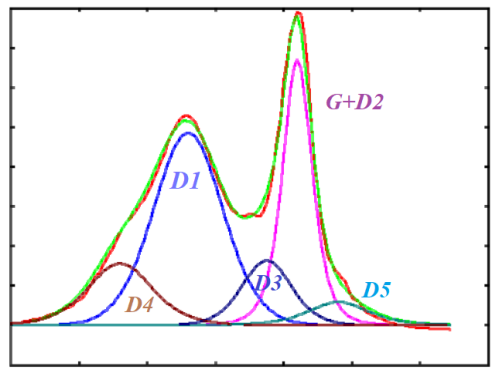

*Fig. 2: The plot here given is obtained by means of 5 q-Gaussians, fitting data from article by Sousa et al., 2020. Arbitrary units are given for both axes. The horizontal axis represents the Raman shift. Here, D2 band has been merged into G band. The green line is representing the fit, and the red line the data. Other colours are giving the q-Gaussian components.*

The fit in Sousa et al., 2020, was made by means of *6 curves*: two Gaussian (D, G) and four Lorentzian functions (D2, D3, D4, D5). Sousa et al. are telling that, "In general, the *deconvolution* of the spectrum can be performed by simple adjustment using only two Lorentzian or two Gaussian functions, which allows for the study of the dispersion behaviour of the D and G Raman bands without a high level of details". But, when the research is aiming to find the "subtle nuances of the spectrum, a 5-band (G, D1, D2, D3, and D4) adjustment is required" (see Sousa et al., 2020, and references therein). The D5 band at 1700 cm$^{-1}$ was inserted to improve the "accuracy of the fit". The Table 2 in Sousa et al. shows general information about the bands.

Claramunt et al., 2015, proposed the study of Raman spectra of graphene oxide and thermally reduced graphene oxide, reporting, in all the recorded spectra, five bands given as D, D′, G, D″, and D*, between 1000 and 1800 cm$^{-1}$. Comparing Sousa et al., 2020, to Claramunt et al., 2015, we have that the notation D means D1, D′ is D2, G remains the same, D" is D3, and D* corresponds to D4.

For the fit of the first-order spectra of the graphene oxide derivatives, Claramunt and coworkers used five functions: two Gaussian and three pseudo-Voigt peaks. In particular it is told that "data were fitted to a sum of five functions using the Origin 8.0 software" (Claramunt et al., 2015). "During the analysis it was found that Gaussian functions better fit D* and D″ bands while, for D, G and D′ bands as pseudo-Voigt functions render the best fit" (Claramunt et al., 2015). The pseudo-Voigt functions given by software Origin are linear combinations of Gaussian and Lorentzian functions.

Yin et al., 2018, have studied the char structure evolution during pyrolysis. In the Figure 4 of their article, we can see that the researchers used four bands (three Lorentzian, G, D1, D4 and one Gaussian, D3). In the article, references about the fit of Raman spectra are given. For chars, (pine needle, PN, and corn stalk, CS), the resulting positions of peaks are: "The band positions of the D1, D3, D4, and G peaks" for the CS char sample "are 1347, 1516, 1212, and 1594 cm$^{-1}$, respectively". "The band positions of the D1, D3, D4, and G peaks" for PN char samples "are 1343, 1514, 1207, 1599 cm$^{-1}$, respectively" (Yin et al., 2018). Yin and coworkers are also giving discussion about the origin of the bands, so that we can tell that D1 band, about 1350 cm$^{-1}$, "is usually related to disordered graphite structures, sp$^2$ hybrid carbon atoms, aromatic rings, edge carbon atoms, and in-plane vibrations with structural defects" (Yin et al., 2018, and references therein). The D3 band, about 1500 cm$^{-1}$, is coming from the "amorphous sp$^2$-bonded carbon, fragments, or functional groups in the disordered structure" (Yin et al., 2018, and references therein), and so on for the other bands. The G band is told representing the "stretching vibration of aromatic rings and the crystal structure of sp$^2$ carbon atoms" (Yin et al., 2018).

In the article by Xu et al., 2020, it is told that Ferrari and Robertson, 2000, "pointed out that the Raman spectrum was the state's vibrational density modified by a coupling coefficient, incorporating various resonances. *Therefore, there was no special reason for choosing a particular function to fit the spectrum*" (Xu et al., 2020). And this is a very interesting observation. "Generally, the symmetric-line fit with two Gaussian lines or two Lorentzian lines were adopted in this method. The most widely used alternative to a Gaussian fit was a Breit−Wigner−Fano (BWF) line for the G band and a Lorentzian fit for the D band" (Xu et al., 2020). The line BWF is asymmetric. For the use of BWF in the Raman spectra of graphene, see please (Hasdeo et al., 2014).

"Other researchers also found that the combination of a Gaussian fit for the D band and a Lorentzian fit for the G band is more suitable for the highly disordered" carbonaceous materials (Xu et al., 2020). We have another interesting further observation proposed by the researchers: "When curve-fitting the Raman spectrum with the two bands, [~1350 cm$^{-1}$, ~1590 cm$^{-1}$] no variable parameter is needed to be fixed" and this is a relevant feature "to establish a common approach" (Xu et al., 2020). But, by using just two bands it is difficult to have an "ideal curve-fitting", and for this reason the multiband fitting has been developed (Xu et al., 2020). Then, it is possible to find used three bands, with an additional band "originally included to improve fitting precision in the deconvolution of Raman spectra. ... In these methods, the Gaussian line was preferred for the band at about 1500 and 1200 cm$^{-1}$, while the Lorentzian line was more used for the band at about 1620 cm$^{-1}$ ... Subsequently, some authors integrated the above methods and developed the four bands method and five bands methods" (Xu et al., 2020). And then, see please again the Table 2 by Xu and coworkers and references given therein.

In Ferrari, (2007), we can find Raman spectroscopy applied to graphene and graphite. "The toll for the simplicity of Raman measurements is paid when it comes to spectral interpretation. The Raman spectra of all carbon systems show only a few prominent features, no matter the final structure, be it a conjugated polymer or a fullerene" (Ferrari, 2007, mentioning Ferrari & Robertson, 2004). The spectra appear quite simple with "a couple of very intense bands in the 1000–2000 cm$^{-1}$ region and few other second-order modulations. However, their shape, intensity and positions allow to distinguish a hard amorphous carbon, from a metallic nanotube, giving as much information as that obtained by a combination of other lengthy and destructive approaches" (Andrea C. Ferrari, 2007, mentioning Ferrari & Robertson, 2004). In the article by Ferrari, it follows a detailed discussion about the main features of the bands, with physical interpretation.

In Yuan and Mayanovic, 2017, an empirical study on Raman peak fitting is given. "Fitting experimentally measured Raman bands with theoretical model profiles is the basic operation for *numerical determination* of Raman peak parameters" (Yuan & Mayanovic, 2017). Yuan and Mayanovic consider various algorithms of "peak fitting", by means of "Gaussian, Lorentzian, Gaussian–Lorentzian, Voigtian, Pearson type IV, and beta profiles". The researchers are proposing an interesting observation: according to the fitting results of the Raman bands, "the fitted peak position, intensity, area ... values of the measured Raman bands *can vary significantly depending upon which peak profile function* is used in the fitting, and the most appropriate fitting profile should be selected depending upon the nature of the Raman bands" (see please the further discussion in Yuan & Mayanovic, 2017). The researchers used PeakFit v.4.11 software package, with other packages, that is LabSpec v.5.78, GRAMS/AI v.9.2 and OriginPro v9.0. However, it is also told that "No significant difference was observed among the peak fitting results by using different software packages, *provided that the same profile* (e.g., Gaussian–Lorentzian) was used" (Yuan & Mayanovic, 2017). This seems saying that comparing the results obtained by means of different profile functions is a sensitive issue. Nevertheless, it is an issue that requires specific attention.

After this short review of literature, we can tell that Lorentzian, Gaussian, Voigt, pseudo-Voigt (linear combination of Gaussian and Lorentzian functions) and Breit−Wigner−Fano (BWF) line are mainly used. However, the problem is not only given by the choice of line shape.

**3. Biasing factors**

In the article by Lünsdorf et al., 2014, we can find mentioned the "bias". The article is a discussion about the problem of comparison of results obtained by means of the Raman spectroscopy of carbonaceous materials, that is, of the RSCM parameters. The comparability "is low as there are at least three major sources of biasing factors. These sources are the spectral curve-fitting procedure, the sample characteristics itself and the experimental design including the used Raman system" (Lünsdorf et al., 2014). Moreover, the researchers have also shown that the "curve-fitting is strongly influenced by individual operator-bias and the degrees of freedom in the model, implying the need for a standardised curve-fitting procedure" (Lünsdorf et al., 2014). And we have also another fundamental fact: "Due to the diversity of components (optics, light detection device, gratings, etc.) and their combinations within the Raman systems, different Raman instruments generally give differing results" (Lünsdorf et al., 2014).

The consequence given by Lünsdorf and co-workers is that to estimate comparable RSCM data, "every Raman instrument needs its own calibration". Therefore, it is necessary to have a series of reference materials.

The Raman spectroscopy is actually a condensed matter of "several biasing factors", arranged in three categories: 1) the first is a bias regarding the curve-fitting, 2) the second is related to the intrinsic nature of CM and 3) the third bias is related to the experimental set-up and related Raman instrument which is used. In the first category we can find the different baseline correction methods, the different line shape functions (Gaussian, Lorentzian, Voigt, etc.), and the different number of bands used as a model. In the second category, we find the sample preparation and heterogeneity. In the third category, we can find for instance the excitation wavelength, the used spectral grating and light detector device.

In Lünsdorf et al., 2014, the above-mentioned issues are analysed. Let us stress that in the article we find also the "Influence of operator's personal fitting strategy". Therefore the researchers used "The spectral processing … performed once with Lorentzian functions only and once with Voigt functions only" (Lünsdorf et al., 2014). In the study by Lünsdorf and co-workers, software Fityk has been used for peak- and curve-fitting (Wojdyr, 2010); "the position and shape of the components are detected automatically by the software. If this is not successful, the components are located manually" (Lünsdorf et al., 2014).

### 4. Fitting the bands

After the discussion about the D and G bands and the related fitting procedures, it is better to add some deeper points, that we can borrow from the article by Robert Meier, 2005, where the author proposed "some crucial issues in curve-fitting vibrational spectra that do not always seem respected". Here in the following some points.

1) In the case of isolated bands, no problem. "Problems arise when there is band overlap" (Meier, 2005). The fitting procedure is used to determine the bands. Curve fitting can also be considered as "modelling", because we need a model for the line shapes.

2) "Curve fitting is finding the best fit to an overlapping band profile starting from a prescribed set of individual bands" (Meier, 2005). And this is what we have found in the previously reported literature.

3) The curve fitting is sometimes also called "deconvolution". However, according to Meier, in the vibrational spectroscopy "deconvolution is generally the process in which instrumental effects are removed from a spectrum" (Meier, 2005). We have that the spectrum is the convolution of the physical line shapes and the broadening caused by instrumentation.

4) A Gauss–Lorentz sum function $S(\omega) = \alpha\, G(\omega) + (1-\alpha)\, L(\omega)$ (*linear combination of functions*), "is not uncommon" (see Meier, 2005, and references therein), "but rarely justified". $\omega$ is the angular frequency. We will discuss the Gaussian - Lorentzian model in a specific section.

5) The line-profile in Raman spectra "is Lorentzian in nature", and Meier is mentioning that there are different manners to arrive at a Lorentzian line. One way refers to a hydrodynamic theory, but more recently we find NMR, optical and mass spectroscopy (Meier, 2005, referring to Marschal & Verdun, 1990). As a simple example, Meier is giving the damped harmonic oscillator.

6) The "shape of individual vibrational bands is thus in essence Lorentzian" (Meier, 2005), but we can have a Gaussian line broadening produced by instrumentation. "Broadening with a Gaussian line profile implies that *one has to multiply the Lorentzian profile with a Gaussian profile*, leading to the so-called Voigt profile" (Meier, 2005). Then, we arrive to the *convolution* of a Gaussian G and a Lorentzian L function.

The Voigt profile (by Woldemar Voigt) is the following probability distribution function:

$$V(x;\sigma,\gamma) = \int_{-\infty}^{\infty} G(x';\sigma) L(x-x';\gamma) dx' \qquad (1)$$

7) "The main difference between a Gauss and a Lorentz profiles is that the latter one has very long tails" (Meier, 2005). And then a "lot of the overall intensity" is buried in tails, "which we generally *do not consider* in a spectrum fit" (Meier, 2005). Really important observation. "The Voigt profile is obviously 'a situation in between'." Very interesting indeed: a q-Gaussian is 'a situation in between' too.

However, see please the detailed discussion given by Meier, 2005.

8) First important remark. Broadening effects lead to convolution. "The regularly used sum of a Lorentz and a Gauss profile has no physical basis, unless ... " see please detailed discussion in Meier, 2005.

9) To fit the spectrum the number of bands is not crucial, "as even two clearly overlapping bands may cause a challenge comparable to an envelope comprising a large number of bands" (Meier, 2005). Taming the bands is a difficult issue. The curve fitting is accomplished by the selection of the wavelength range where there are the bands relevant for "quantifying certain functional groups or constituents of the sample involved" (Meier, 2005). After the spectral range is defined, the base-line determined and then the line shape chosen.

10) Second remark, about the subtraction of background. It is easily performed by software packages. However, there is a problem, because the Lorentz profile has a fat tail and in it we find some intensity of the line; "it is preferred to fit both the envelope and the background simultaneously" (Meier, 2005).

11) About the best fit: "mathematically the best fit (according to least squares fitting) is not a priori the best fit according to the physics of the system. Any good fit has small residuals, but even then various different fits may have acceptably small residuals" (Meier, 2005). An excellent visual result coming from a very low value of the least square fitting does not ensure the physics of the fitting.

12) Let us conclude with the third remark given by Meier, 2005. Have "proper pre-selection of the number of bands, their width and position are highly important".

## 5. Broadening the bands

We have mentioned, about Lünsdorf et al., 2017, the Voigt profile used for Raman bands, and we have also found this profile in the article by Meier, 2005. Let us read his specific words. Meier considers the Lorentzian essence of the line shape, experiencing a "Gaussian line broadening due to instrumental effects or sample characteristics". However, we have also the *Doppler broadening* which is occurring with Gaussian character. "The same holds for the broadening due to the use of a monochromator with rectangular entrance and exit slits. ... Broadening with a Gaussian line profile implies that one must multiply the Lorentzian profile with a Gaussian profile, leading to the so-called Voigt profile" (Meier, 2005). Once again, let us remember that the Voigt profile is the convolution of a Gaussian and a Lorentzian profile.

Linked to the Doppler broadening, we have also the Dicke narrowing (Ciuryło et al., 2001). In the case that the mean free path of an atom is quite smaller than the wavelength of the radiative transition, a narrowing of the line is produced. The atom is changing its speed and direction several times during photonic emission or absorption. In average on the different Doppler states, we find that the atomic line width is narrower than the Doppler width.

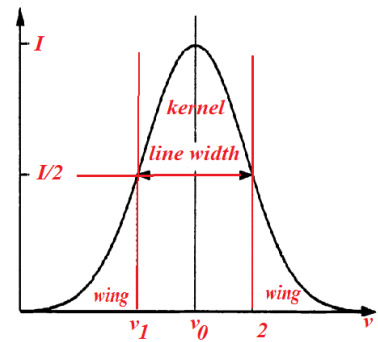

*Fig. 3: Kernel and wings (tails) of a line.*

From book by W. Demtröder, 1985, entitled "Laser spectroscopy: Basic concepts and instrumentation", let us consider width and profile of a spectral line. The spectral lines are never monochromatic. Even in the case of high-resolution interferometers, we observe a spectral distribution $I(v)$ of absorbed or emitted intensity, around the central frequency $v_0$ which is corresponding to an atomic or molecular transition. The central frequency is given by the energy difference between energy levels. The function $I(v)$ in the vicinity of $v_0$ is called the line profile. The frequency interval between the two frequencies $v_1$ and $v_2$, for which $I(v_1) = I(v_2) = I(v_0)/2$ is the Full Width at Half Maximum of the line (FWHM), that is the line-width or half-width of spectral line. The spectral region within the half-width is the "kernel" of the line, the regions outside ($v < v_1$ and $v > v_2$) are the wings (or also the tails).

Let us now consider the natural line shape. An excited atom can emit energy as spontaneous radiation. To have the spectral distribution of a spontaneous emission, according to the difference between energy levels, the simplest model for the excited atomic electron is a classical one with the electron considered as a damped harmonic oscillator (Lorentz model). Being damped, the amplitude $x(t)$ of the oscillation turns out with a frequency lower than the frequency of the undamped case, and the amplitude is decreasing with time. Because the amplitude $x(t)$ is decreasing gradually with time, the frequency of the emitted radiation is not monochromatic, but possesses a frequency distribution related to the function $x(t)$ by a Fourier transformation. The exponentially decreasing amplitude is therefore corresponding to a Fourier transform with a Lorentzian line shape.

In the book by W. Demtröder, 1985, we can also find the spectral profile of an absorption line, derived from atoms at rest. However, being atoms and molecules in thermal motion, we have the additional effect previously mentioned, that is the broadening of the line profile known as Doppler broadening. Then, the Lorentzian-line profile is the natural line, but Doppler effect changes the line.

Let us consider an excited molecule with a velocity in the rest frame of the observer. The central frequency of a molecular emission line has a value in the coordinate system of the molecule, but it is Doppler shifted in the frame of the observer. We know that at thermal equilibrium, the molecules of a gas, for instance, are following a Maxwellian velocity distribution. Since the emitted or absorbed radiant power is proportional to the density of molecules emitting or absorbing in the frequency interval $dv$, the intensity profile of a Doppler-broadened spectral line becomes a Gaussian profile with a full half-width, which is called the Doppler width. But in Demtröder, 1985, it is stressed that "More detailed consideration shows that a Doppler-broadened spectral line cannot be strictly represented by a pure Gaussian profile". The reason is that not all molecules with a definite velocity component are emitting or absorbing at the same frequency. Let us remember that the response of these molecules is represented by a Lorentzian profile, so we arrive at the convolution of Lorentzian and Gaussian profiles, that is at the Voigt profile. To the Doppler effect, we have also to add the broadening due to elastic and inelastic collisions (see details in Demtröder, 1985).

For what concerns the Raman Scattering, it can be considered as an inelastic collision of an incident photon with a molecule in the initial energy level $E_i$. After the collision a photon with lower energy is detected and the molecule is found in a higher-energy level $E_f$ (Stokes radiation). The energy difference $E_f - E_i$ may appear as vibrational, rotational or electronic energy of the molecule. If the photon is scattered by a vibrationally excited molecule, it may gain energy and the scattered photon has a higher frequency. This scattering is called anti-Stokes radiation (see please the detailed discussion in Demtröder, 1985).

## 6. Back to q-Gaussians

Before giving the q-function expression, it is necessary to stress that the term "q-Gaussian" is regarding the deformation of the Gaussian probability distribution function according to Tsallis generalized statistics. It is not a function of the "q-calculus", that is the quantum calculus (about q-calculus, Sparavigna, 2021).

As given by Umarov et al., 2008, the q-Gaussian is:

$$f(x) = Ce_q(-\beta x^2) \quad (2),$$

where $e_q(.)$ is the q-exponential function and $C$ a constant based on $\Gamma$ function. In the exponent, in the discussion proposed here, we use $\beta = 4/\gamma^2$, where $\gamma$ is the parameter of Lorentzian curves. The q-exponential has the expression:

$$exp_q(u) = [1 + (1-q)u]^{1/(1-q)} \quad (3).$$

The plot in the Figure 3 is showing the behaviour of this exponential for different q values. Note that, for q less than one, the function is different from zero on a limited interval.

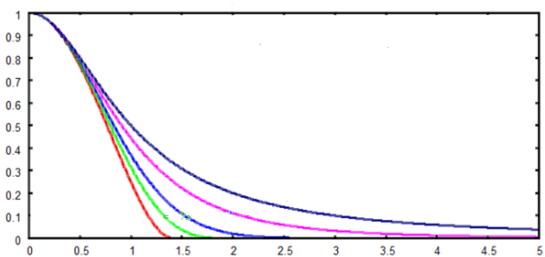

*Fig.4: q- exponential functions in the case of γ=2 (blue, q=2, Lorentzian; pink q=1.5; light blue, q= 1,01 near Gaussian; green, q=0.75; red, q=0.5) For q < 1, the function is different from zero in a limited interval. Being the line symmetric, only the right part of it is given in the figure. The same we will do in the following figures.*

.

## 7. Voigt and Pseudo-Voigt line shapes

As we have seen, the Voigt function is appearing in Demtröder, 1985, in Lünsdorf et al., 2017, and in the discussion by Meier along with the linear combination of Gaussian and Lorentzian functions, which is sometimes also mentioned in the literature about Raman spectroscopy as pseudo-Voigt function.

In Naylor et al., 1995, we can find a publication where Voigt and pseudo-Voigt were compared and differences evidenced. The researchers fitted data by means of a commercial software package, which was giving a "summation of appropriate amounts of Gaussian and Lorentzian profiles" (Naylor et al., 1995). "Data were also subjected to an in-house developed program … based on the Levenberg-Marquardt algorithm. This program uses Voigt line shapes ... to describe the bands" (Naylor et al., 1995). In the curve fitting, the background was considered simultaneously with the spectral bands. For the Voigt function, the given reference is to Armstrong, 1967. This article, entitled "Spectrum line profiles: The Voigt function" is a precious work about the properties of this probability distribution. In the appendix of B. H. Armstrong, 1967, a Fortran program is also given for numerical calculations.

In Naylor et al., no mention to a "pseudo-Voight" approach is given but it is referred to a "LabCalc software program ... describing the Raman bands as sums of a Lorentzian and a Gaussian band shape" (Naylor et al.,1995). We have previously seen that Sadezky and co-workers proposed nine combinations of Lorentzian and Gaussian bands;  then, let us consider in detail LabCalc software, to understand the nature of the combination. The software was made by Failla et al., 1992, who concluded that the "manual deconvolution was as good as computerized" (Naylor et al., 1995).

In Failla and co-workers' article we find that the researchers employed computerized fitting, according to a 'curve fitting' procedure based on a least-square decomposition proposed by Keresztury and Földes, 1990. The program had two alternatives. One approach considers the measured Raman band shape at 1295 cm$^{-1}$ "approximated by the combination of 60% Gaussian and 40% Lorentzian functions) (and Failla et al. are mentioning Keresztury and Földes). The other alternative allows the computer to find the best combination.

In Keresztury and Földes we arrive at the nature of combination. To analyse the spectrum, "one has to determine first the form of the function most closely approaching the observed band shapes. Pure Lorentzian or pure Gaussian functions deviate substantially from the measured band shapes, either near the peaks or on the wings" (Keresztury & Földes. 1990). "By gradually changing the mixing factors in the *Lorentzian-Gaussian sum function*, we found that for all samples of very different crystallinities the 60% Gaussian plus 40% Lorentzian function gave the best fit with the observed spectra" (Keresztury and Földes. 1990). Here the sum function given in the reference:

$$0.6G + 0.4L = I(\nu) = I_0 \left\{ 0.6 \, exp\left[-4 \, ln(2) \left(\frac{\nu_0 - \nu}{w}\right)^2\right] + 0.4 \left[1 + 4\left(\frac{\nu_0 - \nu}{w}\right)^2\right]^{-1} \right\} \quad (4).$$

$I_0, \nu_0, w$ are peak intensity, position and full band width at half height. Let us stress that this line shape is defined as a *combination* of Gaussian and Lorentzian function.

Today, searching for "pseudo-Voigt" on the web we have information mainly regarding software, and find linear combinations like that given above. However, pseudo-Voigts are also those given by Ida et al., 2000, approximating the Voigt functions to 1%. Eq.4 is giving just a sum of functions, but it was successful indeed.

In Zickler et al., 2006, we find "pseudo-Voigt" just mentioned with two references. In the first, Paris et al., there is no "pseudo-Voigt". In the second, (Yamauchi & Kurimoto, 2003), here also there is not mention of a "pseudo-Voigt". The bands "were determined by curve fitting based on the Levenberg-Marquardt method. *A component* of a set of overlapping bands *was regarded as a linear combination* of Gaussian and Lorentzian" (Yamauchi & Kurimoto, 2003). Then, we can conclude that, in literature about Raman spectroscopy, a linear combination of Gaussian and Lorentzian function is seldom also called "pseudo-Voigt", in particular when commercial software is used. In any case, the combination of Gaussian and Lorentzian functions is applied, if it is suitable for, to single components, that is to the single bands of the spectrum, as in the case of Claramunt et al., 2015.

In Ida et al., 2000, we can find the "*extended* pseudo-Voigt function for approximating the Voigt profile". In the article, "The formula of the pseudo-Voigt function expressed by a weighted sum of Gaussian and Lorentzian functions is extended by adding two other types of peak functions in order to improve" the approximation of Voigt profile (Ida et al., 2000). It is also stressed that in the pseudo-Voigt we can use intermediate functions. At Pag. 1312 of Ida et al., 2000, we find told that there are "only a few functions that satisfy both of the following requirements: (i) intermediate shape between Gaussian and Lorentzian; (ii) availability of both the primitive function and its inverse. ... One candidate is an irrational function with the following normalized formula:" (Ida et al., 2000):

$$f(x;\gamma) = \frac{1}{2\gamma}[1 + (x/\gamma)^2]^{-3/2} \quad (5)$$

Let us write again the generic q-exponential: $exp_q(u) = [1 + (1-q)u]^{1/(1-q)}$. We can see immediately that q=5/3. The function that we find in Ida et al., 2000, is the following q-Gaussian:

$$C \, exp_{5/3}(-\beta x^2) = C \, [1 - (1 - 5/3)\beta x^2]^{1/(1-5/3)} \quad (6)$$

Therefore, after all, the q-Gaussian is a component of a "pseudo-Voigt" for sure. But the q-Gaussian is much more. The q-Gaussian can be a computation friendly version of the physical meaningful Voigt line shape too; it is also a statistical distribution, and possesses the two properties mentioned by Ida et al., 2000.

As we can see, Ida et al. are using an irrational function with q=5/3. This value of q is intriguing, because for $q < 5/3$ the variance of the q-Gaussian is finite. When $5/3 \leq q < 2$ it is infinite, and for $2 \leq q < 3$ indefinite (Tsallis et al., 1995).

Keresztury and Földes, 1990, have proposed a linear combination of Gaussian and Lorentzian functions (see Eq.4). Let us call it "pV" and compare it, and other combinations, with the q-Gaussian.

Minimizing: $\Delta^2 = \Sigma(f_{pV} - f_{q-Gau})^2$    (7),

the q parameter is obtained.

Then, let us use:

$$f_{pV} = \left\{ A\, exp\left[-4ln(2)\left(\frac{x}{w}\right)^2\right] + (1-A)\left[1 + 4\left(\frac{x}{w}\right)^2\right]^{-1} \right\} \quad (8)$$

$$f_{q-Gau} = C e_q(-\beta x^2) \quad (9)$$

where $\beta = 4/w^2$

In the following Figure, three combinations are given, with pV in red and q-Gaussian function in green and $w = 1$.

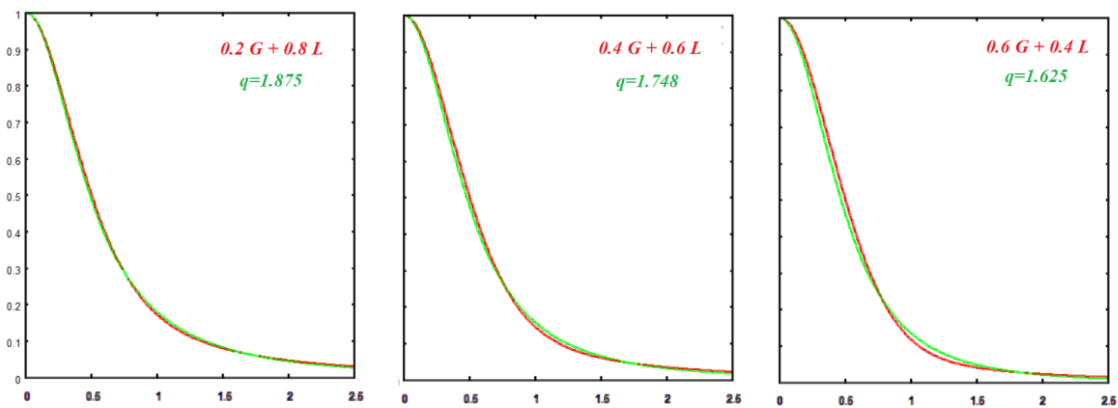

*Fig. 5: Comparing pseudo-Voigt functions ($w = 1$) and q-Gaussians.*

When the weight of the Gaussian component increases further, to fit the pseudo-Voigt with a q-Gaussian function we need to relax the condition $\beta = 4/w^2$ and use a generic $\beta'$ which is determined by the best fit. In the following Figure, we have the combination with A=0.95. The q-Gaussian function needs $\beta'$.

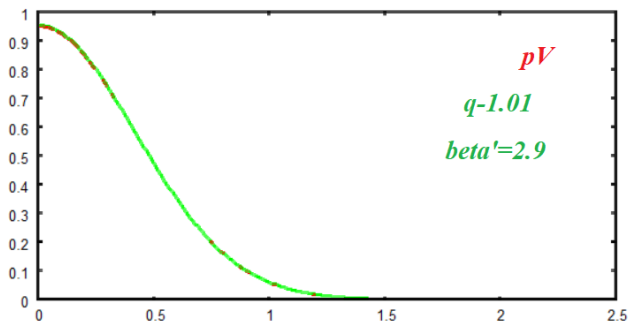

*Fig. 6: Comparison of pseudo-Voigt and q-Gaussian, as previously told.*

In the previous examples, we have assumed the pseudo-Voigt in the form of Eq.8. Let us change it into:

$$f_{pV} = \left\{ A\, exp\left[-4\ln(2)\left(\frac{x}{w_1}\right)^2\right] + (1-A)\left[1+4\left(\frac{x}{w_2}\right)^2\right]^{-1} \right\} \quad (10)$$

with different $w_1, w_2$.

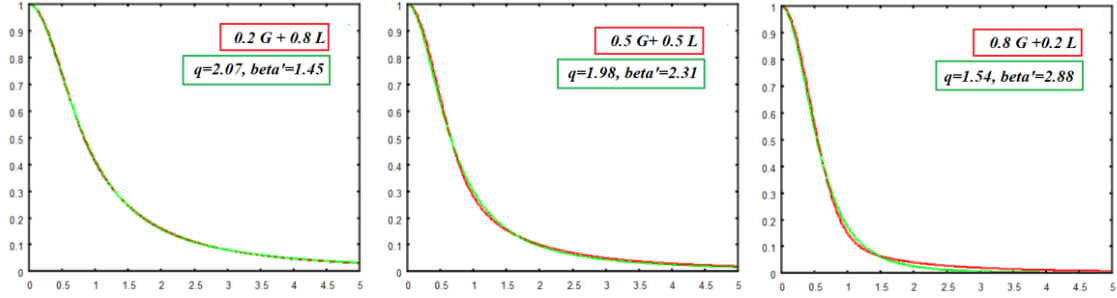

Fig. 7: Comparing pseudo-Voigt functions ($w_1 = 1, w_2 = 2$) and q-Gaussians. In this comparison, the scale factor was $C = 1$. We can see that the best agreement happens when the Lorentzian contrubition is more relevant.

We can conclude that the q-Gaussians can fit the pseudo-Voigt functions, adjusting the values of q, C and $\beta'$, in particular with the best agreement when the role of the Lorentzian component is more relevant. Let us now consider the Voigt function, given as:

$$f_V = C \int_{-\infty}^{+\infty} dx' \left\{ exp\left[-4\ln(2)\left(\frac{x'}{w_1}\right)^2\right] \times \left[1+4\left(\frac{x-x'}{w_2}\right)^2\right]^{-1} \right\} \quad (11)$$

The C constant is given so that the peak of the Voigt function is equal to 1. Let us again compare this function with the q-Gaussian function (9). The results are given in the following images.

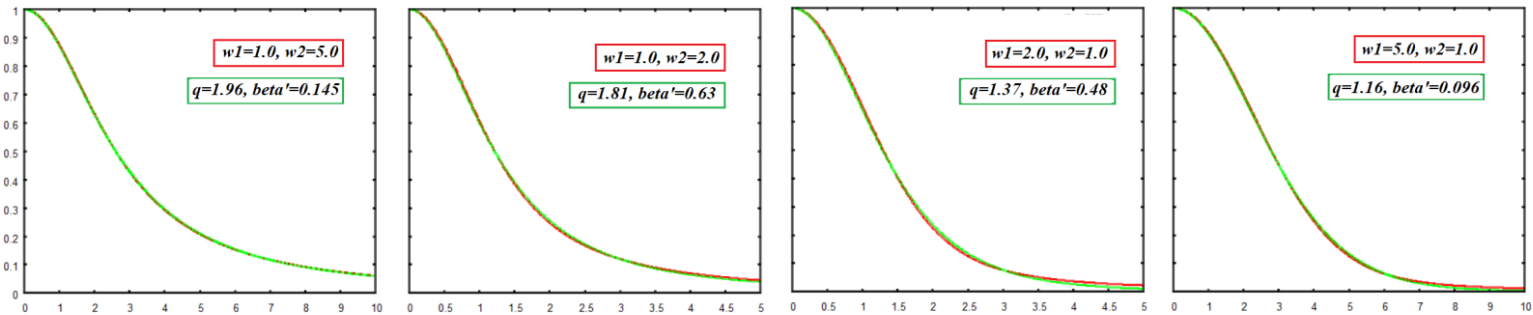

Fig. 8: Comparing Voigt functions and q-Gaussians.

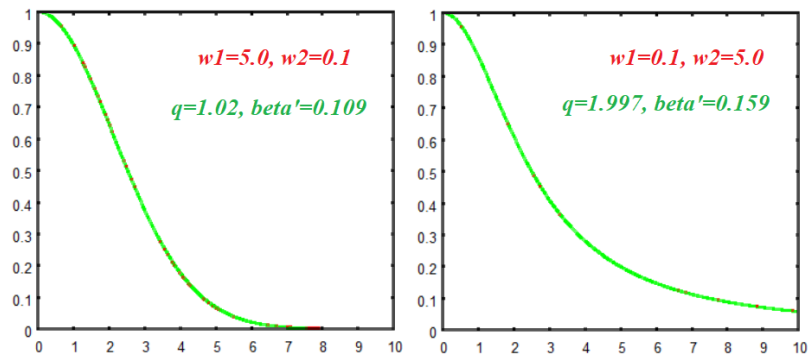

*Fig. 9: Comparing Voigt functions and q-Gaussians.*

The comparison in the case of the Voigt function is even better than in the case of pseudo-Voigt functions. Note that the q-Gaussians can follow the tails of Voigt functions very well.

In Jain et al., 2018, it is given comparisons among Gaussian-Lorentzian sum GLS functions, Gaussian Lorentzian product GLP functions, and Voigt functions. The framework is the peak fitting of X-ray photoelectron spectroscopy (XPS). "Plots of the GLS show that it is a better mathematical representation of a function that is intermediate between a pure Gaussian and a pure Lorentzian. The GLS also appears to be a better approximation of the Voigt function. Because of the more compact nature of the Gaussian, the GLP does not have significant wings" (Jain et al., 2018). Actually, we have seen that Voigt functions have wings with different consistency, and the same is also true for the q-Gaussians, depending on the value of parameter q.

## 8. Tsallis line shape in EPR

In Howarth et al, 2003, it is given a generalization of the line shape, considered as useful in electron paramagnetic resonance (EPR) and possible also in nuclear magnetic resonance (NMR) spectroscopy. And this line shape is defined as "Tsallis lineshape", however it is a q-Gaussian. In the introduction, Howarth and co-workers provide useful information. Here some remarks from their article.

1) Since the initial measurements, the line shapes have been considered as Lorentzian or Gaussian. These functions are simple, with a well-established theoretical basis for their use.

2) A Lorentzian line shape is justified by the linear response theory (Howarth et al. are mentioning R. Kubo, in J. Phys. Soc. Jpn. 9 (1954) 888, but it is Kubo & Tomita, 1954). Following Kubo and Tomita model, it is assumed the magnetization of the sample as oscillating harmonically in external magnetic field. In the presence of a dissipative phenomenon, due to energy transport from the spin system to a heat sink, "the linear response to the electromagnetic (microwave) radiation" produces an exponential decay of the magnetization, and consequently a Lorentzian absorption line "in a continuous-wave EPR (or NMR) absorption measurement" (Howarth et al, 2003).

3) Gaussian line shape is coming from "the inhomogeneity of the resonance field present at each paramagnetic centre. Thus, besides inhomogeneity of the external magnetic field, local stresses, and material imperfections cause all the paramagnetic centres ... to experience a slightly different interaction with neighbouring atoms" (Howarth and co-workers are mentioning Van Vleck, 1948).

4) Then we find statistics. Interactions are giving the "local magnetic field" acting on each centre, and the related resonance frequency. These frequencies can be described statistically. Quoting the central-limit theorem, we arrive to the Gaussian line shape (Howarth et al. mentioning O. Svelto, 1998).

5) In experiments, neither Gaussian nor Lorentzian line shapes are giving satisfactory results. "The effect of rapid changes of the individual centres magnetization vectors can lead to a dynamic averaging of the individual distributions, leading again to a *Lorentzian line-shape at the centre while the wings of the line remain Gaussian*. This effect is known as motional narrowing" (see Howarth et al. and the works by Anderson and Weiss mentioned by them).

6) However, and here we find the proposal by Howarth et al., the "inhomogeneous broadening does not necessarily result in a Gaussian distribution"; therefore we arrive to different line shapes. We find the Voigt line shape function, where Lorentzian and Gaussian shapes are two special cases. Then we have also other generalizations (see references given by Howarth et al.). "A natural extension of the line-shape function is a Levy distribution". But "the major disadvantage of the Levy distributions is that ... there is no analytical expression for these line-shape functions" (Howarth et al., 2003).

7) Extensions of Gaussian and Lorentzian functions are "the distributions introduced by Tsallis". In this manner we arrive to the Tsallis line shape functions (that is q-Gaussians).

8) "In addition to the linewidth parameter, the shape of the line is controlled by one more parameter (q) that determines the contribution of the line wings to the integral intensity" (Howarth et al., 2003).

Then, let us stress that the q-Gaussians, that we are here considering for the Raman line shapes, are the "Tsallis line shapes" proposed in the framework of Kubo-Tomita model.

In R. S. Johal, 1999, we can find an interpretation of Tsallis statistics. In the work by Johal, it is told that P. A. Alemany, 1997, conjectured that the Tsallis formalism is related to a scale invariant statistical mechanics. Therefore, the Tsallis distribution can be obtained "by assuming an ensemble which has replicas of the same system at different scales" (Johal, 1999). Physical examples are the poly-dispersed systems like "polymers, commercial surfactants, colloidal suspensions and critical spin systems on hierarchical lattices". Johal is presenting the Tsallis distribution in the framework of colloidal polydispersity.

## 9. Fitting Raman data

In Tagliaferro et al., 2020, in the Figure 6 of their paper, we can find the Raman spectra for different biochar samples. Let us propose a last comparison, concerning one of the biochar samples given there. The analysis by Tagliaferro et al. is based, for this specific sample, on a four band model (G, D1, D3, D4), and on a new line shape that the authors proposed with the name "GauLor" (see also the discussion in Sparavigna, 2023). As we did in our Figure 2, let us make a fit by means of q-Gaussians. Let us use a four-band model too. The result is given in the following Figure 10. Below the horizontal axis, the colored squares are giving the positions of the peaks as deduced by Tagliaferro et al. The agreement seems being good.

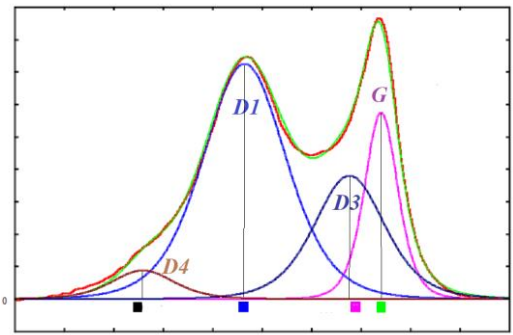

*Fig. 10: The plot here given is obtained by means of 4 q-Gaussians, fitting data from Tagliaferro et al., 2020, regarding a biochar sample. Green (fit) and red (data) curves are almost indistinguishable. The colored squares are giving the position of the bands as deduced by Tagliaferro et al., 2020.*

The D3 band is slightly shifted. However, as it has been previously told by Yuan and Mayanovic, 2017, when the fitting curves are different, differences arise. In the case of the q-Gaussians, the shift is small but exists, being GauLor and q-Gaussian functions two different models of line shape. In any case, further considerations about statistics beyond the models seem being relevant. Let me stress that, in the Figures 2 and 10, I used limited range and data according to the baseline visible in the given references.

In conclusion, we have discussed the q-Gaussian Tsallis line shapes for being applied to the Raman spectral bands. We have seen that the q-Gaussian function is mimicking well the Voigt line-shape. In the case of data from Tagliaferro et al., in the framework of the same four bands model, the agreement is good. Further cases are under study, to generalize the use of q-Gaussians for any Raman bands.